\documentclass[12pt]{article}
\usepackage{amscd,amsmath,amssymb,amstext,amsthm,latexsym}
\newcommand {\al}   {\alpha}       \newcommand {\bt}  {\beta}
\newcommand {\g }   {\gamma}

\newcommand {\s }   {\sigma}

\newcommand {\Lm}   {\Lambda}      
       
\newcommand {\pl}   {\partial}     \newcommand {\nb}  {\nabla}

\newcommand {\CE }  {{\cal E}}      
      \newcommand {\CP}  {{\cal P}}
\newcommand   {\const}{{\sf\,const}}     \newcommand   {\diag}{{\sf\,diag\,}}
\newcommand {\MM}  {{\mathbb M}}   \newcommand {\MR}  {{\mathbb R}}
\newcommand {\MS}  {{\mathbb S}}   
\newcommand   {\ex}{{\sf\,e}}            

\begin{document}
\title     {On homogeneous and isotropic universe}
\author    {M. O. Katanaev
            \thanks{E-mail: katanaev@mi.ras.ru}\\ \\
            \sl Steklov mathematical institute,\\
            \sl ul.~Gubkina, 8, Moscow, 119991, Russia}
\date      {18 June 2015}
\maketitle
\begin{abstract}
We give a simple example of spacetime metric, illustrating that homogeneity
and isotropy of space slices at all moments of time is not obligatory lifted
to a full system of six Killing vector fields in spacetime, thus it cannot be
interpreted as a symmetry of a four dimensional metric. The metric depends
on two arbitrary and independent functions of time. One of these functions is
the usual scale factor. The second function cannot be removed by coordinate
transformations. We prove that it must be equal to zero, if the metric satisfies
Einstein's equations and the matter energy momentum tensor is homogeneous
and isotropic. A new, equivalent, definition of homogeneous and isotropic
spacetime is given.
\end{abstract}

Cosmological models constitute a classic part of general relativity and attract
now great interest because of large amount of observational data. Most of the
cosmological models are based on cosmological principle which requires our
universe to be homogeneous and isotropic [1--5].
\nocite{Friedm22,Friedm24,Lemait33,Robert35,Walker37}
In particular, all space slices corresponding to constant time must be
homogeneous and isotropic.

We consider spacetime $\MM$ with coordinates $\lbrace t,x^\mu\rbrace$,
$\mu=1,2,3$, and metric of Lorentzian signature $(+---)$. We assume that $t$ is
the time coordinate and every section $t=\const$ is spacelike. According to
cosmological principle every section of constant time must be a
three-dimensional space of constant curvature which is homogeneous and
isotropic. Then the usual ansatz for the metric is
\begin{equation}                                                  \label{qnxbgk}
  ds^2=dt^2+a^2\overset{\circ}g_{\mu\nu}(x)dx^\mu dx^\nu.
\end{equation}
where $\overset{\circ}g_{\mu\nu}(x)$ is a negative definite metric on
three-dimensional space slices $\MS$ of constant curvature which does not depend
on time. This metric contains only one arbitrary function $a(t)>0$ which is
called the scale factor. This is the most general form of the homogeneous and
isotropic metric in a suitable coordinate system (see, for example,
\cite{Weinbe72}).

The form of the metric depends on the coordinates chosen on space slices. We
choose stereographic coordinates. Then the metric takes the diagonal form
\begin{equation}                                                  \label{emfree}
  g=\begin{pmatrix} 1 & 0 \\
  0 & \frac{\displaystyle a^2\eta_{\mu\nu}}{\displaystyle \big(1+b_0x^2\big)^2}
  \end{pmatrix},
\end{equation}
where $b_0=-1,0,1$, $\eta_{\mu\nu}:=\diag(---)$ is the negative definite
Euclidean metric, and $x^2:=\eta_{\mu\nu}x^\mu x^\nu\le0$. Since we have chosen
negative definite metric on space slices, the values $b_0=-1,0,1$ describe
constant curvature spaces of positive, zero, and negative curvature,
respectively. For positive and zero curvature spaces, stereographic coordinates
are defined on the whole Euclidean space $x\in\MR^3$. For negative curvature
spaces stereographic coordinates are defined inside the ball $|x^2|<1/b_0$.

Let us make the coordinate transformation $x^\mu\mapsto x^\mu/a$. Then metric
becomes non-diagonal, and the scale factor disappears:
\begin{equation}                                                  \label{emfrei}
  g=\begin{pmatrix}
  1+{\displaystyle\frac{\dot b^2x^2}{4b^2\big(1+bx^2\big)^2}} &
  {\displaystyle\frac{\dot bx_\nu}{2b\big(1+bx^2\big)^2}} \\[2mm]
  {\displaystyle\frac{\dot bx_\mu}{2b\big(1+bx^2\big)^2}} &
  {\displaystyle\frac{\eta_{\mu\nu}}{\big(1+bx^2\big)^2}} \end{pmatrix},
\end{equation}
where
\begin{equation}                                                  \label{qbgsty}
  b(t):=\frac{b_0}{a^2(t)}
\end{equation}
and dot denotes derivative with respect to time.

We see that the metric of a homogeneous and isotropic universe can be
non-diagonal, and may do not contain the scale factor. Moreover, the scalar
curvature of space slices which is proportional to $b(t)$ explicitly depends
on time.

Let us now simply drop non-diagonal terms, add the scale factor, and put
$g_{00}=1$. Then the metric is
\begin{equation}                                                  \label{emfrep}
  g=\begin{pmatrix} 1 & 0 \\
  0 & {\displaystyle\frac{a^2\eta_{\mu\nu}}{\big(1+bx^2\big)^2}} \end{pmatrix}.
\end{equation}
This metric contains two independent arbitrary functions of time: $a(t)>0$
and $b(t)$. It is nondegenerate for all values of $b$ including zero. It is
interesting because allows one to analyse, in a general case, solutions which go
through zeroes $b=0$. If such solutions exist, then the universe can change the
sign of its space slices curvature during evolution.

Metric on space slices $\MS$ of constant time $t=\const$ is
\begin{equation}                                                  \label{qbavsj}
  h_{\mu\nu}=\frac{a^2\eta_{\mu\nu}}{\big(1+bx^2\big)^2},
\end{equation}
depending on time $t$ as a parameter. Straightforward calculations show
that three-dimensional curvature tensor for metric (\ref{qbavsj}) has the form
\begin{equation*}
  \hat R_{\mu\nu\rho\s}=-\frac{4b}{a^2}
  (h_{\mu\rho}h_{\nu\s}-h_{\nu\rho}h_{\mu\s}).
\end{equation*}
Three-dimensional Ricci tensor and scalar curvature are
\begin{equation*}
  \hat R_{\mu\nu}=-\frac{8b}{a^2}h_{\mu\nu},\qquad \hat R=-\frac{24b}{a^2}.
\end{equation*}
We see that that all space slices are spaces of constant curvature. Therefore
there are six independent Killing vector fields $\hat K_i=\hat K_i^\mu\pl_\mu$,
$i=1,\dotsc,6$, which act on space slices. Symmetry transformations on space
slices must be prolongated on the whole spacetime. The usual assumption is that
they act trivially on time coordinate \cite{Weinbe72}. That is Killing vector
fields on spacetime have zero time component:
$K_i:=\lbrace 0,\hat K_i^\mu\rbrace$. By construction, any linear combination of
Killing vector fields $K$ satisfies the Killing equation
\begin{equation}                                                  \label{qbsvwg}
  \nb_\al K_\bt+\nb_\bt K_\al=0,\qquad\al,\bt=0,1,2,3,
\end{equation}
because they do not have time component and time enters space components as a
parameter. The Killing equation can be rewritten for the contravariant
components
\begin{equation}                                                  \label{ekilco}
  g_{\al\g}\pl_\bt K^\g+g_{\bt\g}\pl_\al K^\g+K^\g\pl_\g g_{\al\bt}=0.
\end{equation}
This equation is fulfilled for all moments of time.

The $(\al,\bt)=(0,0)$ component of Killing equation (\ref{ekilco}) for diagonal
metric (\ref{emfrep}) is identically satisfied. The $(\al,\bt)=(0,\mu)$
components reduce to equation
\begin{equation}                                                  \label{qnbcgf}
  \pl_0\hat K^\mu=0.
\end{equation}
The space components $(\al,\bt)=(\mu,\nu)$ of the Killing equation decouple
\begin{equation*}
  h_{\mu\rho}(t,x)\pl_\nu\hat K^\rho+h_{\nu\rho}(t,x)\pl_\mu\hat K^\rho
  +\hat K^\rho\pl_\rho h_{\mu\nu}(t,x)=0.
\end{equation*}
These equations are identically satisfied for all moments of time by
construction.

There is an interesting situation. On one hand, all space slices of metric
(\ref{emfrep}) are homogeneous and isotropic. On the other hand, any
homogeneous
and isotropic metric must have form (\ref{qnxbgk}). The answer is the following.
The whole four-dimensional metric (\ref{emfrep}) is not homogeneous and
isotropic in a sense that Killing equations (\ref{qbsvwg}) are not fulfilled.
Indeed, the six independent Killing vector fields on space slices are
\begin{equation}                                                  \label{qbsghu}
\begin{split}
  \hat K_{0\mu}&=(1+bx^2)\pl_\mu-\frac2bx_\mu x^\nu\pl_\nu,
\\
  \hat K_{\mu\nu}&=x_\mu\pl_\nu-x_\nu\pl_\mu,\qquad \mu,\nu=1,2,3.
\end{split}
\end{equation}
The first three Killing vectors generate translations at the origin of the
coordinate system $x^2=0$, and the last three Killing vectors generate
rotations. We see that the first three Killing vectors explicitly depend on time
through $b(t)$, and Eq.(\ref{qnbcgf}) is not fulfilled. This example shows that
homogeneity and isotropy of space slices does not provide sufficient condition
for the whole four-dimensional metric to be homogeneous and isotropic. The
equivalent definition is the following.

The spacetime $\MM$ is called homogeneous and isotropic if: \newline
1) All sections $\MS$ of constant time $t=\const$ are three-dimensional spaces
of constant curvature;\newline
2) Extrinsic curvature of the embedding $\MS\hookrightarrow\MM$ is homogeneous
and isotropic for all $t$.

Indeed, the first requirement provides the existence of such coordinate system
where metric is block diagonal \cite{Eisenh61}
\begin{equation*}
  g=\begin{pmatrix} 1 & 0 \\ 0 & h_{\mu\nu}(t,x) \end{pmatrix},
\end{equation*}
where $h_{\mu\nu}(t,x)$ is a constant curvature metric for all $t$. For this
metric the extrinsic curvature is \cite{Wald84}
\begin{equation*}
  K_{\mu\nu}=-\frac12\dot h_{\mu\nu}.
\end{equation*}
If it is homogeneous and isotropic, then it must be proportional to the metric,
and we get the differential equation
\begin{equation}                                                  \label{qbncgt}
  \dot h_{\mu\nu}=f h_{\mu\nu},
\end{equation}
where $f(t)$ is an arbitrary sufficiently smooth function of time.

If $f=0$, then nothing should be proved, and the metric has form (\ref{qnxbgk})
for $a=\const$.

Let $f\ne0$. Then we introduce new time coordinate $t\mapsto t'$ defined by the
differential equation
\begin{equation*}
  dt'=f(t)dt.
\end{equation*}
Afterwards equation (\ref{qbncgt}) becomes
\begin{equation*}
  \frac{dh_{\mu\nu}}{dt'}=h_{\mu\nu}.
\end{equation*}
Its general solution is
\begin{equation*}
  h_{\mu\nu}(t',x)=C\ex^{t'}\overset\circ g_{\mu\nu}(x),\qquad C=\const\ne0,
\end{equation*}
where $\overset\circ g_{\mu\nu}(x)$ is a constant curvature metric on $\MS$
which do not depend on time. It implies Eq. (\ref{qnxbgk}).

In general relativity, we assume that metric satisfies Einstein's equations
\begin{equation}                                                  \label{eincos}
  R_{\al\bt}-\frac12g_{\al\bt}R+\frac12g_{\al\bt}\Lm=-\frac12T_{\al\bt},
\end{equation}
where $R_{\al\bt}$ is the Ricci tensor, $R$ is the scalar curvature, and
$\Lm$ is the cosmological constant. The matter energy-momentum tensor is denoted
by $T_{\al\bt}$.

The cosmological principle requires not only the metric but also the
energy-momentum tensor to be homogeneous and isotropic. The most general form
of the homogeneous and isotropic energy-momentum tensor in the coordinate
system defined by Eq.(\ref{emfrep}) is \cite{Weinbe72}
\begin{equation}                                                  \label{engmom}
  T_{\al\bt}=\begin{pmatrix}{\cal E} & 0 \\0& -{\cal P}h_{\mu\nu} \end{pmatrix},
\end{equation}
where $\CE(t)$ and $\CP(t)$ are the matter energy density and pressure.

One can easily calculate the Einstein tensor for metric (\ref{emfrep}). The
off-diagonal component is
\begin{equation*}
  R_{0\mu}=-\frac{4\dot b x_\mu}{(1+bx^2)^2}.
\end{equation*}
To satisfy Einstein's equations (\ref{eincos}) we must put
\begin{equation*}
  \dot b\quad\Leftrightarrow\quad b=\const
\end{equation*}
because all other terms are diagonal. Thus for homogeneous and isotropic matter
we return to the original metric (\ref{emfree}) on the equations of motion.

Metric (\ref{emfrep}) describes spacetime which has homogeneous and isotropic
space slices. We have shown that this important property is not sufficient for
describing homogeneous and isotropic universe. The reason is that three of
the six Killing vectors on space slices do depend on time, and their lift to the
whole spacetime does not satisfy the four-dimensional Killing equations. The
sufficient condition for the metric to be homogeneous and isotropic is (i) all
space slices must be spaces of constant curvature and (ii) time derivative of
the spacial part of the metric must be homogeneous and isotropic (in the
coordinate system described above). Fortunately, metric of type (\ref{emfrep})
seems to be excluded by Einstein's equations, thus there is no reason to worry
too much about this dilemma.

This work is supported by the Russian Science Foundation under grant
14-50-00005.

\end{document}